# Precision Measurement of M1 Optical Clock Transition in Ni$^{12+}$

Shaolong Chen[1*], Zhiqiang Zhou[1,2*], Jiguang Li[3§], Tingxian Zhang[4], Chengbin Li[1], Tingyun Shi[1], Yao Huang[1], Kelin Gao[1#], Hua Guan[1,5#]

[1]State Key Laboratory of Magnetic Resonance and Atomic and Molecular Physics, Innovation Academy for Precision Measurement Science and Technology, Chinese Academy of Sciences, Wuhan 430071, China.
[2]University of Chinese Academy of Sciences, Beijing 100049, China
[3]Institute of Applied Physics and Computational Mathematics, Beijing 100088, China
[4]School of Science, Lanzhou University of Technology, Lanzhou; 730050, China
[5]Wuhan Institute of Quantum Technology, Wuhan 430206, China

[*] These Authors contributed equally to this work
[§/#] **Authors to whom correspondence should be addressed:** li_jiguang@iapcm.ac.cn, klgao@apm.ac.cn, guanhua@apm.ac.cn

**Key words:** Highly charged ions, Atomic optical clock, Precision spectroscopy, Electron beam ion trap

**PACS numbers:** 06.30.Ft, 32.30.Jc, 32.10.Fn, 39.30.+w

## Abstract

Highly charged ions (HCIs) have drawn significant interest in quantum metrology and in search for new physics. Among these, Ni$^{12+}$ is considered as one of the most promising candidates for the next generation of HCI optical clocks, due to its two E1-forbidden transitions M1 and E2, which occur in the visible spectral range. In this work, we used the Shanghai-Wuhan Electron Beam Ion Trap to perform a high-precision measurement of the M1 transition wavelength. Our approach involved an improved calibration scheme for the spectra, utilizing auxiliary Ar$^+$ lines for calibration and correction. Our final measured result of the M1 transition wavelength demonstrates a five-fold improvement in accuracy compared to our previous findings, reaching the sub-picometer level accuracy. In combination with our rigorous atomic-structure calculations to capture the electron correlations and relativistic effects, the quantum electrodynamic (QED) corrections were extracted. Moreover, comparing with an estimate of the one-electron QED contributions by using the GRASP2018 package, we found that the present experimental accuracy is high enough for testing the higher-order QED corrections for such a complex system with four electrons in the *p* subshell.

## Introduction

Atomic optical clocks have emerged as the most precise devices ever created, making them ideal for fundamental research in physics, including experiments that test fundamental physical theories, search for dark matter, and applications in atomic timekeeping, navigation, and geodesy [1-5]. In recent years, various atomic clocks have reached a high level of accuracy of E-18 even E-19 [6-11]. However, with such high precision, the influence of atomic structural properties on further improvements of accuracy is becoming more and more prominent. This is where highly charged ions (HCIs) have an advantage over neutral atoms or singly ionized ions since they are compact and thus insensitive to external electric and magnetic field perturbations [18, 20]. In addition, the transition frequencies in HCIs are more sensitive to variations in fundamental physical constants, such as the fine structure constant. These unique properties make HCIs promising candidates for the next generation atomic optical clocks [12-15].

Since the conception of HCI atomic clocks was postulated [15-17], a diverse array of HCIs have been discovered as suitable candidates for atomic clock [12,18–26]. Meanwhile, numerous challenges in the development of HCI clocks have gradually been overcome [14,25,27,28]. For instance, PTB in Germany has achieved the $Ar^{13+}$ clock with an uncertainty of $2.2\times10^{-17}$, which provides a proof-of-principle demonstration of an HCI optical clock. The associated instability of $2.6\times10^{-14}\tau^{-1/2}$ is limited mainly by the natural linewidth of the optical transition. Additionally, HCI optical clocks based on other competitive species, such as $Ni^{12+}$, $Pd^{12+}$, $Pr^{9+}$, $Nd^{9+}$, are also being proposed and constructed [13,22,30].

$Ni^{12+}$ possesses a relatively simple energy level structure (see Fig. 1), with two visible forbidden optical transitions: the magnetic dipole (M1) transition at 511 nm and the electric quadrupole (E2) transition at 498 nm. This makes it a highly advantageous system among many HCI candidates, with a target uncertainty of E-19 or even smaller [13]. The M1 transition of $Ni^{12+}$ serves as both a candidate optical clock transition to demonstrate the HCI clock and a logic transition for detecting the E2 clock transition, as previously implemented in the $Al^+$ optical clock [11,31]. The precision of the $Ni^{12+}$ M1 clock transition is currently determined only theoretically to the sub-nanometer level, which falls short of the desired accuracy for a clock transition probe. Locating a spectral line with an Hz width within this range is akin to searching for a needle in a haystack. Therefore, before investigating optical transitions of natural width, it is necessary to determine accurate wavelength through line broadening techniques.

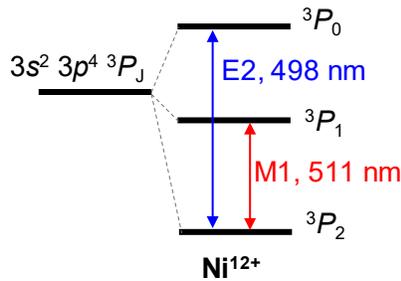

FIG. 1. Energy level diagram of $Ni^{12+}$.

To achieve relatively high precision in identify the forbidden transition lines of HCIs, various methods can be employed, for instance, analyzing the spectrum of solar corona or other hot astrophysical objects [32], or observing the spectra lines from HCI ions generated in a laboratory setting using tokamaks [33–37] and EBIT [19,22,24,38–40]. In recent years, many preliminary wavelength searching experiments of HCI candidate clock transitions have been carried out in EBIT [19], such as the M1 transition of $Ar^{13+}$ with uncertainty of 2E-7 [38], the E2 transition of $Pt^{9+}$ with uncertainty of 2E-6 [22], and the M1 transition of $^{127}I^{7+}$ with uncertainty of 3E-6 [41]. In fact, the M1 transition of $Ni^{12+}$ was first observed in the inner solar corona [32]. Our team previously measured the M1 transition wavelength of 511.570(2) nm using the Shanghai-Wuhan Electron Beam Ion Trap (SW-EBIT), with an accuracy of 4E-6 [24]. The primary limitation of our measurement was the calibration error arising from the incomplete overlap between the calibration lamp and the ion. To address this, here we utilize a position-non-critical calibration scheme and auxiliary Ar II lines for calibration and correction, thus reducing the calibration error. As a result, we have obtained the most accurate M1 transition wavelength in $Ni^{12+}$, with a fractional uncertainty of 8E-7. Our finding will provide a narrower wavelength range for further probing the $Ni^{12+}$ clock transitions.

**Experimental Setup**

The SW-EBIT [42] was used in the experiment to generate $Ni^{12+}$ ions through electron bombardment of nickel dichloride, which contained $^{58}Ni$ atoms with a natural abundance of 68%. To focus on the electron beam, a strong magnetic field of 0.15 T was established by a superconducting Helmholtz coil. The electron beam had an energy of 400 eV and a beam current of 7-10 mA. Because of the space charge effect, the electron beam created a potential trap in its radial direction to capture the ions, while the axial confinement was achieved by applying an additional electrostatic field.

The trapped $Ni^{12+}$ ions were electronically excited by electron impact to a manifold of states, and the resulting spontaneous fluorescence was detected by the spectrometer, a Czerny-Turner spectrograph (Andor Kymera 328i) equipped with an Electron Multiplying Charge-Coupled Device (EMCCD, Andor Newton 970). The spectrometer had a focal length of 328 mm, and an 1800l/mm holographic grating was used. The entrance slit of the spectrometer was set to 30 μm and was immediately followed by an iris to achieve the best signal-to-noise ratio and resolution. The EMCCD had an imaging area of 25.6×3.2 $mm^2$ with 1600×200 active pixels, and in the vertical direction, only the central 140 pixels were binned to reduce the coma and other nonparaxial aberrations that caused deviation. In this imaging system, the HCI cloud was imaged first by two 200 mm convex lenses, where the distance between the HCI cloud and the first lens was equal to the focal length of the lens, as shown in Fig. 2. Thus, the real image of the HCI cloud was behind the second lens at the same distance where the slit of width 100 μm was placed. Behind the slit, another lens was used to take a secondary image of the ion cloud, and then at the entrance of the spectrometer, a real image of the slit and HCI cloud can be seen. To calibrate the system, the calibration light was reflected to the slit using a moveable reflector and diffused by a moveable diffuser. The reflector and diffuser were only slid into place only when the calibration line was exposed, so the slit can serve as a light source for both the HCI cloud and calibration line, making them overlap. This calibration procedure was based on the $Ar^{13+}$ spectral measurement scheme [38]. With this scheme, the position of the calibration lamp was not important, and the calibration error was thus reduced.

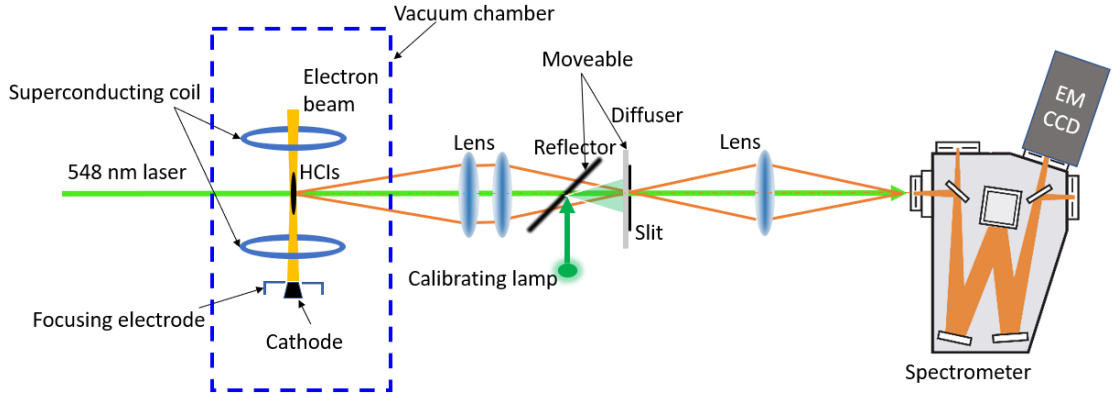

FIG. 2. $Ni^{12+}$ optical transition measurement and calibration scheme.

In our experiment, we sequentially exposed the calibration light and HCI ion fluorescence for 1 minute and 10 minutes respectively. This approach might have caused temperature drift, potentially affecting wavelength measurements. To mitigate this issue, we insulated the affected spectrometer to minimize temperature variations. We evaluated the rate of temperature variation and found it to be only about 0.1 K/h, which corresponds to a wavelength change of 1 pm/h. This improvement is significant compared to the 2 pm/h deviation without insulation. Each acquisition cycle lasted approximately 11 minutes, allowing us to estimate that the deviation caused by temperature variations is 0.17 pm or lower per exposure. Since we performed spectral measurements over several days with dozens of exposures, we can average out the offsets caused by temperature fluctuations during the period. Therefore, these offsets are negligible.

To calibrate the spectrometer's wavelength, we used a Pt-Ne hollow cathode lamp and selected eight well-established lines around 511 nm as calibration lines. These lines had uncertainties smaller than 0.1 pm, and we fitted the recorded profiles with a Gaussian curve. We plotted the positions identified on the EMCCD against their recommended wavelengths and fitted them using a quadratic polynomial as the dispersion function.

Figure 3 depicts the typical exposure of the 511 nm spectral line and a calibration line. A single exposure of the $Ni^{12+}$ optical transition takes 600 s, while the calibration line takes only 60 s. The profiles were obtained by accumulating the images vertically, and a Gaussian fit was used to determine the FWHM of the single 511 nm line exposure, which is about 60 pm. For the calibration line, the FWHM is about 50 pm. The limited resolution of the spectrometer and the Doppler broadening due to ionic motion are the main causes of the observed FWHM of the 511nm line profile. Both effects can be well described by the Gaussian function, and therefore, the convolution of the two should yield a Gaussian profile. The Doppler broadening of the calibration line is negligible in the observed profile; thus, the 50 pm linewidth can be considered as the resolution of the spectrometer. The Doppler broadening of the 511 nm line can be estimated to be 33 pm, corresponding to a temperature of HCIs about 37 eV. The ion temperature can be further reduced by lowering the axial potential well and the electron beam current, which can decrease the spectral linewidth even further.

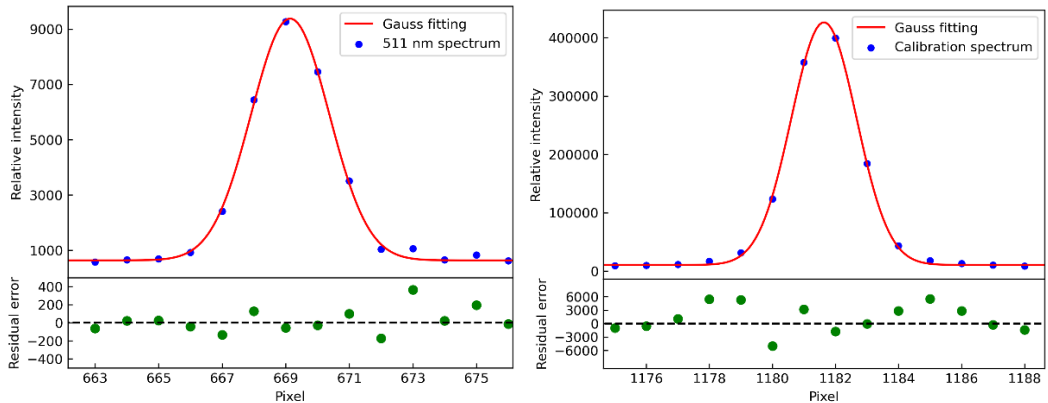

FIG. 3. A Gaussian fit and residual are shown for a single spectrum measurement (on the left) and one of the calibration lines (on the right), with the FWHM of 2.85 pixels and 2.43 pixels for the two profiles, respectively. These values correspond to 60 pm and 50 pm, respectively.

To calibrate each 511 nm measurement, we utilized two sets of calibration exposures taken before and after the 511 nm exposure. By averaging the results of these two sets, we mitigated the impact of system drift. We acquired dozens of measurements over several days of intermittent acquisition to obtain the statistical distribution of the 511 nm spectral line, with the statistical uncertainty of 0.21 pm. We also carefully evaluated the systematic uncertainty and identified the calibration system as the primary source of error. This system encompasses calibration line uncertainty, calibration optical path, and dispersion function.

In our experiment, we determined the calibration line uncertainty to be 0.04 pm based on the NIST database. The error introduced by the calibration optical path can be attributed to the incorrect positioning of the calibration lamp. To address this, we moved the calibration lamp off-axis and measured the wavelength difference between the 20 mm off-axis and on-axis positions, resulting in a shift of 1.19(38) pm/20 mm. Additionally, we moved the diffuser position 50 mm away from the slit and observed no significant shift in wavelength, i.e., a shift of approximately 0.06(9) pm/50 mm. Upon comprehensive analysis, we concluded that the maximum position error of the calibration lamp in the optical path building is around 2 mm, equivalent to a corresponding wavelength error of 0.1 pm. Regarding the error introduced by the dispersion function fitting, we used second- and third-order polynomials to fit the calibration lines and observed no significant deviation in the fitting. We determined the standard deviation of dispersion function fitting to be 0.2 pm. A typical calibration process was shown in Fig. 4, where suitable calibration lines in the observed range were selected based on their richness and isolated positions. Then, the calibration lines were fitted with a second-order polynomial to obtain the wavelength of 511 nm spectral in the fitting curve.

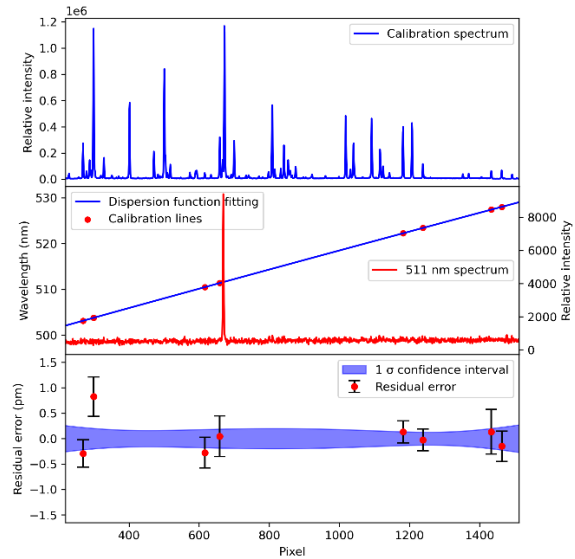

FIG. 4. Top: calibration spectrum of the Pt-Ne hollow cathode lamp. Middle: the pixel positions of the calibration lines with their recommended wavelengths (red points) [43] fitted with a second-order polynomial (blue) as the dispersion function. The red color shows the spectrum of $Ni^{12+}$ acquired over 10 minutes to indicate the position of the pixel relative to the calibration lines. Bottom: the residuals with 1σ confidence band of the dispersion function fit.

The sample used in the experiment was composed of natural abundance Ni. The only stable isotope of Ni with a non-zero nuclear spin is $^{61}$Ni, which has a natural abundance of 1.1%. Thus, the proportion of hyperfine structure in the sample was 1.1%. An estimation of the hyperfine structure revealed a maximum hyperfine splitting of 16 pm [44]. Since we cannot resolve the hyperfine splitting, it only causes the broadening of the spectral lines.

Another effect to consider is the isotope shift in the transition frequency $\nu_{AA'}$ between two isotopes $A$ and $A'$, which can be expressed in terms of two terms:

$$\delta\nu_{AA'} = K\mu_{AA'} + F\delta\langle r^2\rangle_{AA'}.$$

Here $\mu_{AA'} = M_A^{-1} - M_{A'}^{-1}$ with $M_A$ and $M_{A'}$ being the masses of the two isotopes, and $\delta\langle r^2\rangle_{AA'} = \langle r^2\rangle_A - \langle r^2\rangle_{A'}$ represents the difference between their nuclear root-mean-square charge radii. The electronic parameters in the isotope shift, including the mass-shift factor $K$ and field-shift factor $F$, were estimated by using the GRASP2018 package [45-46]. Based on this estimation, the isotopic shifts of $^{60}$Ni, $^{61}$Ni, $^{62}$Ni, and $^{64}$Ni were found to be approximately -0.18 pm, -0.26 pm, -0.34 pm, and -0.49 pm, respectively, relative to $^{58}$Ni. To account for this, we superimposed the Gaussian lines of intensity 26%, 1%, 4%, and 1%, and deviation of 0.18 pm, 0.26 pm, 0.34 pm, and 0.49 pm, respectively, onto another Gaussian line of intensity of 68% with no deviation. This resulted in a composite Gaussian line with a deviation of 0.06 pm, from which we estimated the isotopic shift to be 0.06 pm, with an uncertainty of 0.06 pm.

Additional error terms, such as the Stark effect resulting from the space charge of the electron beam and the second-order Zeeman effect caused by the 0.15 T magnetic field, were discovered to be significantly smaller than the current measurement uncertainty. Therefore, these effects can be safely disregarded.

To further enhance the measurement's reliability, we measured several lines of Ar$^+$ using the same parameters as Ni$^{12+}$. Under identical experimental conditions and adopting the same calibration lines, the calibration system errors for Ar$^+$ and Ni$^{12+}$ ought to be identical. Moreover, we performed alternating measurements of the Ar$^+$ and Ni$^{12+}$ spectra to significantly reduce the drifts of system parameters. We compared the measured Ar$^+$ spectrum with the recommended wavelengths from the NIST database and obtained the difference in wavelength $\Delta\lambda$ together with the statistical error. As depicted in Fig. 5, for the four measured Ar$^+$ lines, $\Delta\lambda$ is within ±0.6 pm. We accounted for the variance resulting from the calibration system and used the standard deviation of the statistical distribution of the Ar$^+$ lines to estimate the calibration system's error. The 1σ uncertainty is 0.2 pm. Furthermore, we considered an average shift of approximately -0.13 pm in the measured $Ar^+$ lines as the uncertainty in the calibration system.

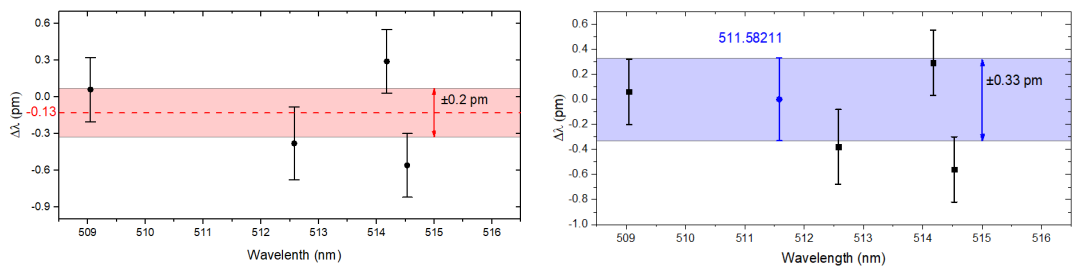

FIG. 5. The left panel illustrates the deviation of the four measured spectral lines of $Ar^+$ compared to the NIST database values, with 1σ statistical error. The right panel compares the position of the measured $Ni^{12+}$ 511 nm spectral line with the $Ar^+$ spectrum after considering a calibration error of 0.33 pm.

Therefore, we chose the frequency shift and uncertainty of the $Ar^+$ measurement as the calibration error for the 511 nm spectral measurement of $Ni^{12+}$ and took the "root sum of squares" to give the overall calibration system error, which is 0.33 pm, as marked on the right panel of Fig. 5. The calibration error estimated with the $Ar^+$ line measurement is comparable to that of the previous analysis of the experimental system, and the estimated uncertainty contains some systematic errors that we may have missed. Thus, we took the uncertainty of 0.33 pm as the final calibration error for the $Ni^{12+}$ measurement. Table I is the error budget, outlining the primary sources of error and their contributions to the wavelength measurements.

Table I. Error budget.

| Source of error | Shift (pm) | Error (pm) |
|---|---|---|
| Line centroid determination | 511582.05 | 0.21 |
| Calibration system | / | 0.33 |
| Isotope shift | 0.06 | 0.06 |
| Stark shift | / | <0.01 |
| $2^{nd}$-order Zeeman effect | / | <0.01 |
| Total | 511582.11 | 0.40 |

**Atomic structure calculation**

Since the high level of accuracy was achieved in the present experiment, a comparison of the fine structure splitting with a rigorous atomic structure calculation could test the QED effects [47-52]. To this end, we carried out calculation on the M1 transition energy by using the multi-configuration Dirac-Hartree-Fock (MCDHF) method and the GRASP2018 package [45-46]. The dominant contributions to the fine structure splittings are the relativistic inter-electronic interaction (electron correlation), which includes the Coulomb interaction

$$\sum_{i<j} \frac{1}{r_{ij}}$$

and the Breit interaction

$$-\sum_{i<j} \frac{1}{r_{ij}} [(\boldsymbol{\alpha_i} \cdot \boldsymbol{\alpha_j}) + \frac{(\boldsymbol{\alpha_i} \cdot \boldsymbol{r_{ij}})(\boldsymbol{\alpha_j} \cdot \boldsymbol{r_{ij}})}{r_{ij}^2}].$$

In the framework of the MCDHF method, the atomic state wave function $|\Gamma J M_J P\rangle$ can be accounted for by systematic expansions of the configuration state functions (CSFs) with the same total angular momentum $J$, its z-component $M_J$, and the parity $P$:

$$|\Gamma J M_J P\rangle = \sum_{\alpha}^{N_{\text{CSF}}} c_\alpha |\gamma_\alpha J M_J P\rangle,$$

where $|\Gamma J M_J P\rangle$ is the atomic state function (ASF) concerned, $N_{\text{CSF}}$ is the number of CSFs, $\{c_\alpha\}$ are the expansion coefficients, and $\Gamma$ and $\gamma_\alpha$ represent other quantum numbers labeling the ASF and each CSF, respectively

For the present case, there are two holes in the $1s^2 2s^2 2p^6 3s^2 3p^4$ ground configuration, so the electron correlation effects are expected to be larger than those systems with a single hole in the outermost shell such as F-like ions. To describe the electron correlation effects, we included single, double, triple, and quadrupole excitations of electrons occupied in the 2s, 2p, 3s, and 3p orbitals to the active orbital set that consists of the spectroscopic and correlation orbitals. The spectroscopic orbitals are those occupied orbitals in the ground configuration $1s^2 2s^2 2p^6 3s^2 3p^4$, which were optimized together with $1s^2 2s^2 p^6 3s 3p^4 3d$ and $1s^2 2s^2 2p^6 3s 3p^2 3d^2$ in the self-consistency field (SCF) calculations. The correlation orbitals are augmented layer by layer, and each layer is composed of orbitals with different angular symmetries up to $n \leq 11$, $l \leq 6$. The correlation orbitals in the added layer were varied adapting to the configuration space generated by single (S) and restricted double (rD) excitations of electrons occupied in $1s^2 2s^2 2p^6 3s^2 3p^4$ and single-double (SD) excitations from $3s 3p^4 3d$ and $3s^2 3p^2 3d^2$. The restricted double excitation means that only one electron in the $n \leq 2$ shells can be excited. Note that in the MCDHF calculations the off-diagonal matrix elements between the SD-excitation CSFs were fully neglected. With respect to the small effects from the high-$l$ orbitals, the number of $g$, $h$ and $i$ correlation orbitals was limited to five, two, and two. The CSFs generated by the triple and quadrupole excitations from $2s^2 2p^6 3s^2 3p^4$ to the first layer of correlation orbitals were included in evaluating the relativistic configuration interaction as well as the Breit interaction. In addition, the frequency-dependent Breit interaction was also estimated at the level of Dirac-Hartree-Fock.

The calculated M1 transition energy is displayed in Table II, revealing that the electron correlations and the Breit interaction play dominant roles, contributing 1.2% and 2.4%, respectively, to this transition energy. The current theoretical result, excluding QED correction but considering the frequency-dependent Breit interaction, stands at 19511 cm$^{-1}$. This value exhibits a 30.758 cm$^{-1}$ less when compared to the experimental value of 19541.758(15) cm$^{-1}$. Based on a comprehensive comparison of the fine structure separations in the ground state of F-like ions between the MCDHF and *ab initio* QED calculations, as concluded in Refs. [50, 51], it becomes evident that the GRASP package can effectively handle the frequency-dependent Breit interaction. Therefore, the discrepancy between the current experimental and non-QED theoretical values can be interpreted as the experimental QED contribution. We roughly evaluated the first-order QED corrections in the fine structure constant α, specifically the one-loop radiative diagrams (the vacuum polarization and self-energy) using the GRASP2018 package [53]. Our estimated result of 40 cm$^{-1}$ is roughly in agreement with the aforementioned experimental QED value, and the discrepancy can further be attributed to higher-order QED effects like screened and two-loop radiative diagrams, among others. It is important to note that the present experimental precision is sufficient to test these higher-order QED effects. However, performing an *ab-initio* QED calculation for such an atomic system with two holes in the valence shell is challenging.

TABLE II. Various theoretical contributions to the M1 transition wavelength in Ni$^{12+}$.

| Effect | Contribution | |
|---|---|---|
| | cm$^{-1}$ | nm, air |
| DHF | 19741 | 506.42 |
| Electron correlations | 242 | -6.13 |
| Breit | -460 | 11.78 |
| Frequency-dependent Breit | -12 | 0.31 |
| QED | 40 | -1.05 |
| Total | 19551 | 511.33 |

Table III lists a comparison of experimental and theoretical M1 transition wavelength in Ni$^{12+}$. One can see that our measured wavelength is significantly more accurate than the theoretical value, with a three orders of magnitude improvement. Additionally, it represents a five-fold improvement compared to our previous measurement [24]. Furthermore, a previous discrepancy of roughly 10 pm between the prior results and the NIST database values has been resolved due to the calibration issue. In our previous work [24], we used a Kr lamp and chose the calibration line at approximately 511 nm. However, this line had a significantly low intensity and was susceptible to stray background interference from nearby high-intensity lines, consequently impacting the accuracy of the entire measurement. We confirmed this by employing the Kr lamp, which exhibited a deviation of roughly 10 pm in both the $Ni^{12+}$ and $Ar^+$ spectra, implying that the calibration line for the selected Kr lamp had undergone a shift. In the current measurement, we applied a Pt-Ne lamp endowed with an increased number of calibration lines and higher intensity. Throughout the measurement process, we adhered to the identical calibration spectra, calibration lamps, calibration lines, calibration optical paths, and spectrometer parameters for both the $Ni^{12+}$ and $Ar^+$ lines. Consequently, numerous systematic errors were found to be similar. Henceforth, the $Ar^+$ line serves as the best auxiliary calibration line, guaranteeing the precision and reliability of the $Ni^{12}$ measurements.

Table III. Comparison of experimental and theoretical M1 transition wavelength in $Ni^{12+}$.

| Year | Type | Wavelength cm$^{-1}$ | Wavelength nm, air | Reference |
|---|---|---|---|---|
| 2023 | Expt. | 19541.758(15) | 511.58211(40) | This work |
| 2021 | Expt. | 19542.2(1) | 511.570(2) | [24] |
| 2018 | NIST[a] | 19541.8 | 511.581 | [43] |
| 2023 | Theor. | 19551(10) | 511.33(26) | This work |
| 2021 | Theor. | 19540(21) | 511.7(6) | [24] |
| 2018 | Theor. | 19560(20) | 511.1(5) | [13] |
| 2018 | Theor. | 19534 | 511.8 | [54] |

[a]Ritz wavelength.

In terms of theory, our calculations have demonstrated that the interaction between electrons is one of the most significant factors. Additionally, the QED effects have played a vital role in the system. The main computational uncertainty in theory was estimated to be about 10 cm$^{-1}$, which arises from the remained fractional electron correlation and the high-order QED effects. It is important to emphasize that accurately modeling and comprehending these effects is essential in studying systems of highly charged ions.

**Conclusions**

We have successfully measured the M1 optical transition wavelength of $Ni^{12+}$ at 511 nm, reaching the sub-pm level with a fractional uncertainty of 8E-7, which represents the most accurate wavelength measurement for highly charged nickel ions to date. Nonetheless, the accuracy of our measurement could be improved even further by adopting a higher resolution spectrometer and performing evaporative cooling to reduce the ion temperature. The use of a high-resolution spectrometer would enable us to acquire better calibration, whereas lower ion temperatures would allow us to obtain more stable and narrower spectral lines. Moreover, the experimental technique applied in this research is applicable to other M1 transition measurements. Furthermore, after taking into account electron correlations and QED effects, we have presented the most precise calculation of the M1 transition. To achieve even greater accuracy, the advancement of higher-order QED theory is warranted..

Our improved measurement accuracy has narrowed down the range of optical transition probing for subsequent development of the $Ni^{12+}$ optical clock. $Ni^{12+}$ presents an opportunity for the creation of two optical clocks within one system by exploiting the two forbidden transitions. The first clock transition uses 511 nm, and the second clock transition uses 498 nm together with the 511 nm line as the logic transition for the logic spectrum, improving efficiency and maintaining authenticity. Theoretical calculations predict that the precision and stability of the $Ni^{12+}$ clock could reach the E-19 level or even higher. Furthermore, nickel comprises four stable isotopes, namely $^{58}Ni$, $^{60}Ni$, $^{62}Ni$, and $^{64}Ni$, and the use of high-precision spectroscopic measurements of 511 nm and 498 nm could aid in the search for hypothetical fifth forces and test the nuclear QED recoil effects through isotope shift measurements.

**Acknowledgments**

The authors thank Shiyong Liang and Jun Xiao for early contributions to the experimental apparatus. and Zongchao Yan for helpful discussions and careful reading of this manuscript. This work was supported jointly by the National Natural Science Foundation of China (Grant Nos.11934014, 12004392, 12121004, 11874090), the Natural Science Foundation of Hubei Province (Grant No. 2022CFA013), the CAS Youth Innovation Promotion Association (Grant Nos. Y201963, Y2022099), and the CAS Project for Young Scientists in Basic Research (Grant No. YSBR-055).